# SAFEMYRIDES: Application of Decentralized Control Edge-Computing to Ridesharing Monitoring Services

Samaa Elnagar[1]*, Manoj A. Thomas[2], Kweku-Muata Osei-Bryson[3]
[1]Howard University, [2]University Of Sydney, [3]Virginia Commonwealth University

**Abstract**
Edge computing is changing the face of many industries and services. Common edge computing models offload computing which is prone to security risks and privacy violation. However, advances in deep learning enabled Internet of Things (IoTs) to take decisions and run cognitive tasks locally. This research introduces a decentralized-control edge model where most computation and decisions are moved to the IoT level. The model aims at decreasing communication to the edge which in return enhances *efficiency* and decreases *latency*. The model also avoids data transfer which raises *security* and *privacy* risks. To examine the model, we developed *SAFEMYRIDES,* a scene-aware ridesharing monitoring system where smart phones are detecting violations at the runtime. Current real-time monitoring systems are costly and require continuous network connectivity. The system uses optimized deep learning that run locally on IoTs to detect violations in ridesharing and record violation incidences. The system would enhance safety and security in ridesharing without violating *privacy*.

**Keywords:** Decentralized control; Edge Computing; Deep Learning; Internet of Things; Ridesharing; Monitoring, Privacy.

**Highlights**
- A shift is needed towards decentralized control for IoTs to decrease computational costs and maintain privacy.
- Technologies such as Deep learning for IoTs and decision support could be used to detect ridesharing violations locally on smartphones.
- Monitoring ridesharing sessions is technically and socially challenging.
- The system doesn't record the entire ridesharing session, but it only records violation incidents.
- The system enhances the ridesharing services by ensuring *privacy, responsiveness, accuracy, and efficiency.*

# 1    Introduction

On the one hand edge computing led to huge reduction in communication cost and speedup computation (T. Wang, 2020). However, with tens of IoTs trying to connect single edge node, computation bottlenecks and delays are still problems. Moreover, edge computing didn't fully address security and privacy issues (T. Wang, 2020). On the other hand, little research has focused on moving and computation and decentralizing decisions to IoTs (Elnagar & Thomas, 2020). What inspire decentralized edge-models that smartphones' specifications as the Internet of Things (IoT) are continuously enhanced while the cost is fairly decreasing (El Khaddar & Boulmalf, 2017). Moreover, optimized deep learning models could run on the limited processing, batteries, and memory of IoTs to provide stand-alone decision-making and cognitive tasks (Elnagar & Thomas, 2020). Decentralizing control and federating the computation at the IoT level will minimize security, privacy and latency risks that emerge from transferring data to other computation nodes (Li & Taeihagh).

One of the IoT applications that can't contemplate privacy and decision latency is the ridesharing service where safety is a significant concern (Feeney, 2015). Some countries in Europe and developing parts of the world report ridesharing to be comparatively unsafe (Chaudhry, El-Amine, & Shakshuki, 2018), which suggests the dire need of imposing additional safety measures to protect both the driver and the passenger (Hong, 2017). From the social and governance aspects, recording the entire rideshare session violates passengers' privacy (Anderson, 2016; Z. Lee, Chan, Balaji, & Chong, 2016). Technical challenges include storing and processing ridesharing monitoring sessions, and latency in detecting violations. Moreover, cellular communication in developing countries, where crime rates are higher, is limited (Dillahunt, Kameswaran, Li, & Rosenblat, 2017). In attempts to solve the latency and cloud dependency issues, edge computing-based monitoring solutions focused on increasing the transmission bandwidth, optimizing workloads, and increasing the number of edge processing nodes (T. Wang, 2020). However, these solutions not only necessitate costly special infrastructure but also impose security and privacy risks. In high traffic, the latency issue would be doubled due to the increased processing workload dedicated to each edge unit (Meshram, Choudhary, & Velaga, 2020; Zhang, Wang, Lu, Liu, & Shi, 2019). Therefore, edge computing-based solutions partially solve the latency issue, but the privacy, and network dependency issues are not entirely solved.

This research proposes a decentralized-control edge model that aims to fill the *efficiency, responsiveness, and privacy* gaps in current edge computing systems by moving most of the computation and decision making to the IoT level. With application to ridesharing monitoring, the model minimizes cost represented in network dependency, costly infrastructure, or sacrificing latency. So, we introduce a local decentralized scene-aware monitoring system for ridesharing services using the driver's smartphones. To ensure passenger privacy, the system record only violation incidences instead of recording the entire rideshare session. The system applies state-of-the-art deep learning optimized models to smartphones to detect ridesharing violations. To ensure *security*, The system encrypts the detected violations before submitting them to ridesharing agents. The implementation of the system exhibits superior detection latency among current solutions.

So, the broad research problem is shifting the gear towards decentralized edge computing and empowering the decision making of IoTs. The narrow research focus is implementing *SAFEMYRIDES*, a decentralized control to in ridesharing monitoring to address the safety concerns in ridesharing sessions.

*SAFEMYRIDES* is motivated by the lack of real-time monitoring systems in ridesharing, the high cost (connectivity cost, infrastructure cost, and processing cost), and the preach of privacy of the current solutions (continuous monitoring of passengers). The system also could be applied in developing countries where high computational resources and mobile internet connections are limited.

In the following sections, we review prior research on ridesharing monitoring. Then, an overview of ridesharing violations, followed by the methodology and description of the proposed system. The architecture of the system is described and evaluated experimentally and using illustrative scenarios. Finally, future work and limitations are provided.

## 2 Literature Review

### 2.1 Current Research in Edge Computing

Current edge computing research is dedicated to improving media processing and computation latency and accuracy on the edge computational units. Video Analytics in Public Safety (VAPS) is one of the most successful edge computing applications (Zhang et al., 2019). However, due to the high real-time requirements and communication overhead, it is hard to implement VAPS at scale or in developing countries. Long, Cao, Jiang, and Zhang (2017) developed a cooperative video processing scheme using an edge computing framework to enable cooperative processing on resource-abundant mobile devices for delay-sensitive multimedia IoT tasks. In the research, several edge nodes cooperate for enhanced media task preprocessing performance. Yi et al. (2017) presents the LAVEA system, an edge computing platform, which offloads computation between clients and edge nodes, and collaborates with nearby edge nodes. The research aims to provide low-latency video analytics at places closer to the users to minimize the response time.

(L. Wang, Zhang, Li, Zhong, & Shi, 2019) designed MobileEdge, a three-task offloading system that shares computing tasks to nearby on-road computational units. The system target achieving optimal task scheduling for collaborative computing. The system used *Tensorflow lite (TF lite)* that is used in our research. The results show that MobileEdge significantly reduces the application response latency. However, on-road computational units exist only in certain areas in developed countries, Moreover, collaborative computing with other computational units is prone to serious security threats and privacy violations. Simaraly, *DeepDecision*, a mobile deep learning framework for edge video analytics, was developed (Ran, Chen, Zhu, Liu, & Chen, 2018). The framework ties together computationally weak front-end devices with more powerful back-end helpers to allow deep learning to choose local or remote execution and determine an optimal offload strategy. Yet, the system assumes a well-defined shared computation infrastructure.

### 2.2 Previous Ridesharing Monitoring Systems

Monitoring ridesharing sessions has been a controversial issue. According to Ulrich Beck's concept of risk society (Beck, Lash, & Wynne, 1992), surveillance risks mitigate the safety avails offered. Therefore, using monitoring systems in ridesharing would enhance safety and consequently increase user satisfaction. Y. S. Lee (2018) conducted a study on *Lateral Surveillance* in Singapore. They found that contrary to the belief that individuals are always skeptical about surveillance, the respondents, on average, reported a positive perception of lateral surveillance. Previous ridesharing monitoring systems include the Safe-ShareRide system developed by L. Liu, Zhang, Qiao, and Shi (2018). The system can detect violations happening in the vehicle by detecting a predefined list of seeking help phrases. The system uses speech recognition for verbal help detection. Then, the system starts recording videos and analyze them on the nearest computational points. However, depending on the seeking help phrases for violation detection is not typical in most cases, especially for passengers under influence.

Kumar, Mukherjee, and Singh (2017) used smartphones for monitoring roads through sensing the road surface and conditions from a moving vehicle using fuzzy logic-based road surface roughness classification. The system collects accelerations of users; processes the information and communication them to a central server. Then, a georeferenced database is updated based on collected information and visualization on Google® maps. To obtain a broader picture of the current status of ridesharing monitoring, Table 1 provides a comparison between previous systems and the proposed system in terms of methodology and service quality issues.

Table 1: A Comparison Between Pervious Edge computing Systems and The Proposed System.

| Research | Methodology | Computation Level | Latency | Privacy | Efficiency |
|---|---|---|---|---|---|
| Kumar et al. (2017) | Fuzzy logic-based | Central Server | - | - | yes |
| Ran et al. (2018) | Mobile deep learning, | Local or remote execution | yes | - | yes |
| Long et al. (2017), Yi et al. (2017), Zhang et al. (2019) | Edge computing | Nearby edge nodes, or Shared IoT | yes | - | - |
| L. Liu et al. (2018) | Speech recognition for verbal help detection | At IoT (speech recognition), edge unit (video) | yes | - | - |
| Proposed Approach | Tensorflow lite, Optimized deep learning | At the IoT (smartphone) level | yes | yes | yes |

## 2.3 Violations in Ridesharing

According to most ridesharing companies, there is an agreeable code of conduct to be followed by both passengers and drivers. A common code of conduct includes (Armant & Brown, 2020; BILL): Avoid sitting next to the driver on the passenger seat. Avoiding any physical interaction with the driver. Physical interaction with the driver includes violent actions, seductions, and arguments. Moreover, the driver cannot sit next to the passenger in the back seats (Macmurdo, 2015). A Driver must wear a seatbelt all the time. The code of conduct also illegalizes the use and hold of weapons during ridesharing sessions. In addition, the code of conduct condemns any sexual conduct between the driver and passenger (Lyft, 2019). In the design of the ridesharing monitoring system, we trained the system's deep learning model to detect these types of violations.

## 3  Conceptual Framework

The gap in literature could be summarized that current edge solutions are oriented towards optimizing the computation and media processing on the cloud or nearby edge nodes (Cheng, Bao, Zarifis, & Mou, 2019; Z. Lee et al., 2016). Also, there is an inclination towards using edge analytics for media processing where multiple edge units are collaborating to decrease the latency. On the other hand, solutions that share the computation are inefficient requiring costly infrastructure that is hard to be implemented in developing countries and rural areas. In addition sharing computation is violating the *privacy* and *security* of user's data (Cheng, Su, Luo & Cai, 2021). Finally, none of the edge solutions are trying to move computation locally to the IoT level to decrease network reliance and ensure privacy.

On the other hand, the ridesharing monitoring issues discussed in the literature section could be summarized as *the lack of privacy, safety, and late response*. These issues have affected customer satisfaction and trust. Most edge computing research avoided moving computation to the IoT level because of the limited computational resources of IoT. Therefore, ensuring system *efficiency* in terms of battery, memory, and storage consumption in our research is crucial. So, the proposed model aims at enhancing *efficiency, (Functionality), privacy, and responsiveness* as key factors to enhance IoTs applications performance such as ridesharing monitoring services. The description of each factor is discussed below:

*Privacy*: gauges the degree to which the IoT application protects customer information and only with their consent, disseminate it to other business activities (Hizam & Ahmed, 2020). Since mobile devices are accessible to sensitive information including biometric features, privacy has been an essential factor in assessing electronic applications (Wolfinbarger & Gilly, 2003; Zeithaml, Parasuraman, & Malhotra, 2002). In SAFEMYRIDES, the ridesharing session is recorded based on passenger consent and only violation incidence is recorded. Moreover, recorded media (photos or videos) will be encrypted before they are saved.

*Efficiency:* measures the resources and ease with which a system is accessed and used (Zeithaml et al., 2002). The decentralized-control edge approach aims to consume fewer resources in terms of less network communication that consume less battery and data transfer cost. The approach also uses DL optimization techniques such as quantization and compression of DL networks. Quantization maximize memory savings during the inference process, and compression of DL networks decreases the size of the network to few megabytes, makes it efficient to be used in low storage devices (Jacob et al., 2018). Moreover, efficiency will be evaluated by the resources consumed such as memory, storage, and battery consumed.

*Responsiveness:* is one of the most important service quality dimensions for human-computer-interaction (HCI) (Ho & Lee, 2007). Responsiveness is how the system delivers the results of an operation to users in a timely and organized manner. Latency is directly correlated with responsiveness, the delayed is the response, the less responsiveness in the system. Therefore, the proposed approach will be evaluating the average latency as a measure of *Responsiveness*.

*Accuracy:* there are different definitions of accuracy. In the context of *SAFEMYRIDES*, accuracy is defined as the ability of the system to detect violations correctly out of all violation incidences. The confusion matrix of precision and recall is also a principal part of calculating accuracy.

## 4 Methodology

In this research, we follow a design science approach (Hevner, March, Park, & Ram, 2004). The decentralized-control edge computing approach is considered a model artifact, and the ridesharing monitoring system (*SAFEMYRIDES*) is considered a framework artifact (Peffers, Rothenberger, Tuunanen, & Vaezi, 2012). The framework artifact is implemented to show functionality and test performance. The system is evaluated as instantiation using synthetic or real-world situations to evaluate the detection *accuracy*, *latency, privacy, and consumed resources*. So, the evaluation is performed using *controlled experiments* or the study of the artifact in a controlled environment for qualities. Next, an overview of decentralized-control edge computing approach is presented.

### 4.1 Decentralized Control Edge Model

Edge computing was built upon the concept of computation offloading where end devices and IoTs perform data collection and sensing (Huda & Moh, 2022) as shown in the right side of Fig.1. Complex processing and real time analysis are performed on the edge units or on the cloud. However, running machine learning and decision-making algorithms on IoTs is now an attainable task fueled by the increase in computational resources are significant decrease in computational cost (Mocnej et al., 2021). Complex computation tasks such as deep learning inference could run on IoTs with regular consumption of resources. The decentralized control edge model is proposed on the left side of fig.1. The model limits the computation on the edge or the cloud side to limit traffic and computational overheads.

The model moves most of computation and decision-making to the end device level to allow decentralized control (task onloading). The role of edge units and/or cloud platforms in the decentralized-control model is archiving/logging events of the IoT. So, IoTs connect edge units or the cloud platforms for saving logs or upgrading the processing of tasks. The model not only decreases network traffic between end devices and edge units but also, allow a real-time response to IoT sensed events. Thus, this model focuses on IoTs *Task Onloading* and IoT-edge communication reduction. However, since deep learning models are complex and resources exhaustive, deep learning tasks should be optimized first before they could be applied to IoTs as discussed in the next sections

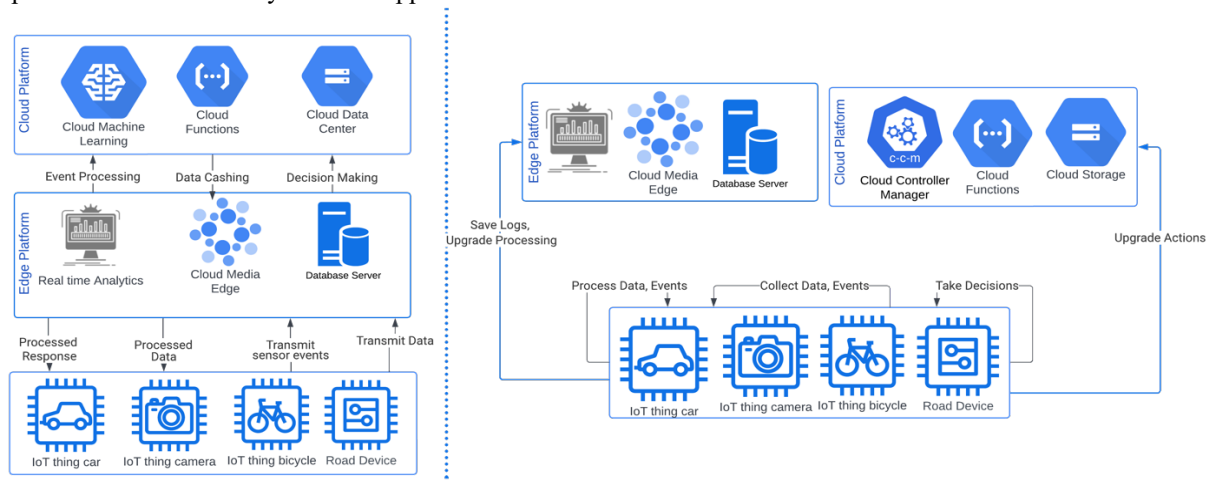

Fig. 1: Decentralized Control Edge Model

## 4.2 Deep Learning Networks Optimization

*DL* models are known for their high computational cost and complexity, so they mostly run on servers on the cloud (Wu et al., 2018). On the other hand, *end devices (e.g.,* IoTs*)* and *edge devices* (e.g., routers) are often battery-powered, have limited memory, and processing to store and process data. Applying *DL* models to *IoT* is challenging; DLNs must be optimized and compressed to fit IoTs limited computational sources. In addition, the optimization of the IoTs themselves is also necessary in terms of memory and hardware optimization (Synced, 2017).

Compression is one of the main optimization techniques that aim to reduce the massive size of *DL* networks. One of the popular compression methods is the *Pruning* technique that eliminates the connections between neurons to directly reduce the feature map width and shrink the network size. Quantization is another optimization technique that aims at compacting the number of bits required to store the *DLN* weights, usually from 64 bit to 8 bits (Han, & Dally, 2015).

## 4.3 SAFEMYRIDES Design

The proposed system is an on-site decision support system for monitoring ridesharing vehicles, where the driver's smartphone can detect the violation and decides the proper action. We trained a Mobile SSD network to detect violation incidences. Then, the trained Mobile SSD is optimized to be deployed to smartphones. The design of the proposed system is shown in Fig. 2 and the components of the system are discussed below in detail.

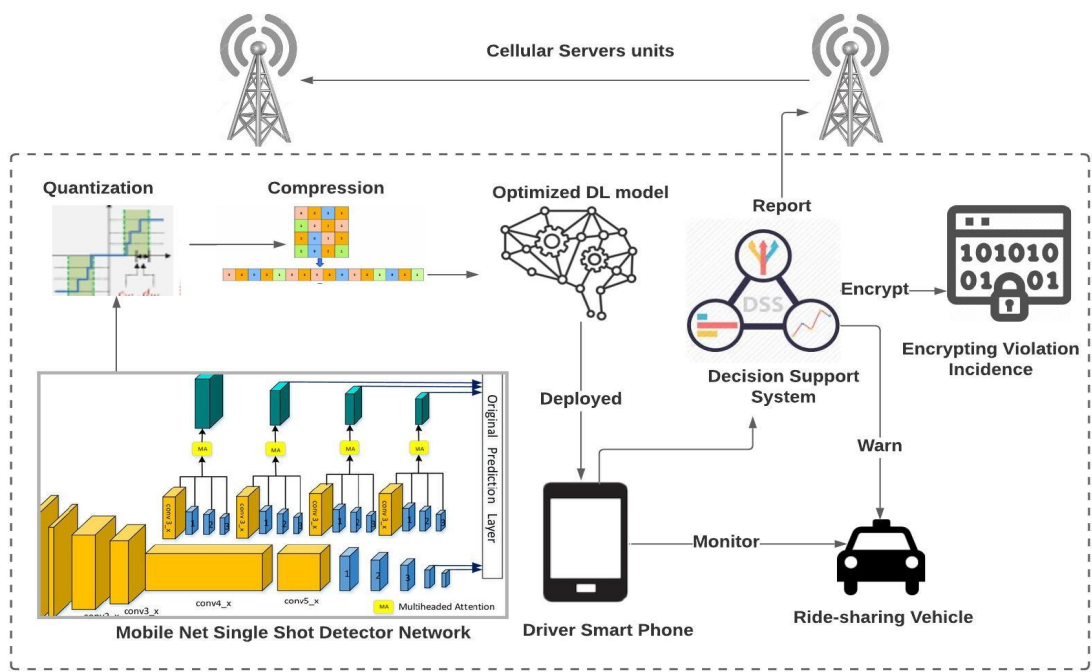

Fig. 2: Decision Support System for Monitoring Ridesharing Vehicles.

### 4.3.1 Optimized Deep Learning (DL) Model

Usually, Deep learning models (DLMs) consist of many connected layers resulting in millions of parameters and weight. So, running DLMs tends to be a resource-intensive process in terms of energy consumption and memory accesses and hardware accelerators. To apply a DLM to IoTs and small devices such as smartphones, DLMs must be compressed and quantized. Also, the DLM running on IoTs are read-only means that these compressed models cannot be retrained or edited. We used a Mobile Single Shot Detector (Mobile SSD) which is based on a feed-forward convolutional network that produces a fixed-size collection of bounding boxes and scores for the presence of object class instances in those boxes. SSD is designed to be independent of the base network, and so it can run on top of any base network such as VGG, YOLO, MobileNet (W. Liu et al., 2016).

The DLM is quantized where parameters are represented in 8 bits instead of 32 bits to accelerate the inference process. The DLM will have a post-training quantization which is the most used form of quantization. In this approach, quantization takes place only after the model has finished training. Moreover, the network is compressed to be 10x less in size of the trained network following the method presented by (Singh, Manikandan, & Namboodiri, 2019).

### 4.3.2 The Ridesharing Smart Phone

Recently smartphones' specifications are enhancing rapidly at a very reasonable cost. The average processing power of most smartphones currently could host an optimized DLM and run inference smoothly. For the system to run smoothly, the smartphone should have a minimum RAM (not less than 4GB) and an acceptable resolution camera according to TensorFlow Guide[1].

### 4.3.3 Decision Support System (DSS)

The DSS is built inside the smartphone. It is responsible for acting in case of detected violation. If a violation is detected by the Mobile SSD, a warning is voiced to passengers/drivers. If violations continue to be detected, the DSS will record the incidence, encrypt it, and save the incidence in a hidden folder in the ridesharing's smartphone. Then, the recorded incidence is sent to the ridesharing agent and attached with encrypted media.

### 4.3.4 Encrypting and Transferring Violation Incidence

For security, images or videos that document the violation incidences are encrypted and saved in the ridesharing's smartphone storage to ensure passenger/driver privacy. In addition, encryption ensures secure transfer to the ridesharing agent. Otherwise, violation evidence could be deleted before it is reported to the ridesharing agent. On the other hand, encryption will ensure security if the media is shared with nearby road units. So, passengers' information will be kept confidential without driver intervention. Moreover, passenger/driver data is secured in case the phone is stolen or hijacked.

### 4.3.5 Transferring Violation Evidence using The Cellular Server Units (CSU)

After the violation incidence is securely recorded and encrypted, it will be uploaded to the nearest cellular server units. In case of no cellular connection, Violation incidences could be uploaded to the nearest on-road units in smart cities infrastructure or sent in text messages. CSUs could be the driver cellular network or the networks that exist on the roadsides in developed countries. So, mostly CSUs are the nearest access point where the DSS could upload incident videos even if no cellular internet is turned off on the phone.

### 4.3.6 The Ridesharing Vehicle

Information about the ridesharing vehicle is saved in the ridesharing agent database. If the violation requires legal authorities' interventions, the ridesharing vehicle could be found easily. In addition, the ridesharing agent requires registered vehicles to have a valid title, proper condition, and valid insurance of the driver.

## 5 Implementation

The implementation process is divided into two steps. The first step is the training and testing of the Mobile SSD network used in detecting violations. In this step, we collect, annotate, and pre-process relevant ridesharing violation samples. Then, after the network is fully trained, it is optimized (compressed and quantized) to be easily embedded in IoT(smartphone). The second section is implementing the DSS system that runs locally on the driver's smartphone. The DSS act when the Mobile SSD detects violations. The DSS actions include sending warnings, encrypting violation incidences, and reporting them to the ridesharing agent. In the following sections, the implementation of each section is discussed in detail.

### 5.1 Implementing the Trained Mobile SSD Network

#### 5.1.1 The Training Dataset

There is no ready to use dataset that contains images of ridesharing violations incidents. The training and testing samples used to train the Mobile SSD model are retrieved from web searches, ridesharing agents' articles, and public ridesharing violations videos. The violation images were a collection of the five-standard code of conduct rules such as interacting physically with the driver as specified earlier. The images collection was taken during the day and at night. We also included normal images where there is no violation.

---

[1] https://www.tensorflow.org/lite/guide

### 5.1.2 Annotation and Preprocessing:

Images were annotated in three classes: driver, passenger, and violation. Annotation was performed using Roboflow[2]. Dataset was divided into 90% training, 6 % validation, and 4 % testing. Images were annotated and saved in TFRECORD format. Every class in the image was surrounded by a bounding box representing a driver (*pink* box), passenger (*green* box), and violation (*purple* box), as shown in fig. 3. Noted that all passengers are annotated if the image has more than a passenger. Image preprocessing included *resizing* pictures to 416 x 416, *Auto-Orientation,* and *Auto-Adjust of Contrast* through *contrast stretching* (Munteanu & Lazarescu, 1999).

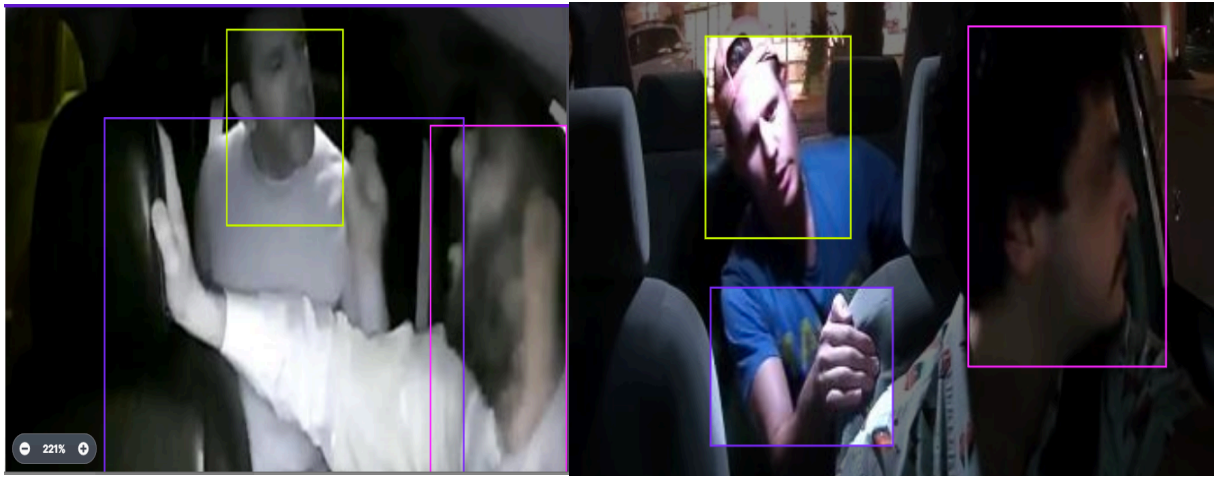

Fig. 3: Annotated Training, Validation Images Where Each Class is surrounded with A Bounding Box.

### 5.1.3 Hardware and Software

The Training was performed on the cloud Google Collaboraty using Jupiter notebooks. To accelerate the training process, we used one GPU. The network was trained using TensorFlow 1 and a Keras version less than 2.0 as recommend for custom models training settings[3]. The ridesharing DSS was created using Android studio 4.1.2 and SDK 28. The device used for running and debugging is a Mac-book pro with intel core i5 2.5 GHz processor, 8 GB Ram. The build and debug were performed on a Google PIXEL AVD before the system was deployed to a smartphone.

### 5.1.4 Optimized MobileNet Single Shot Detector Network

We trained a MobileNet Single Shot Detector (v2) (ssd_mobilenet_v2) that is popular for performing object detection tasks on IoT devices. *ssd_mobilenet_v2* is optimized for lightweight inference, enabling it to perform well natively on compute-constrained mobile and embedded devices. The Mobile SSD is quantized to parameters size of 8 bit. Quantization consumes less memory bandwidth. Fetching numbers in the 8-bit format from RAM requires only 25% of the bandwidth of the standard 32-bit format. On the other hand, quantizing neural networks results in 2x to 4x speedup during inference. We didn't train the network from scratch; rather, we transferred learning from a pre-trained model with a COCO dataset. COCO dataset training weights were uploaded to the training network in the form of .h5 format. Also, we used the same configuration for MSCOCO Dataset as shown below in table 2.

Table 2: Trained Network Configuration.

| Parameter | Value |
|---|---|
| The number of layers | 6 |
| Minimum network scale | 0.2 |
| Max Network scale | 0.95 |
| Number of steps | 100000 |
| Activation Function | RELU_6 |

---

[2] https://app.roboflow.com/
[3] Model Customization

After the network was trained, we ran an inference test, and then we compressed the network and quantized it to 8 bit. Then, the trained network was converted into *TFLITE* format. The 8 bit quantization format is used to run inference on IoTs such as smartphone devices. The average time spent per training step is `global_step/sec:4.96517.` So, the entire training session duration lasted 13.7921 hours.

### 5.2 Implementing the SAFEMYRIDES Decision Support System

#### 5.2.1 Detection Response

The DSS was developed in Java for Android using Android Studio IDE. The DSS primary interface runs a scene view stream of the smartphone camera. The stream is not recorded. Otherwise, the detection runs on the fly using the smartphone camera running scene. The system has access to both the front and rear cameras. When a violation is detected with confidence level greater than 80 %, the system speaks out "*Violation Detected*" using the *Text to sound* android library. If another violation is detected, the system saves the current scene into an encrypted image file and stores it in a hidden folder in the ridesharing's phone. The average inference time of the system is 28 ms for detecting the three classes of (passenger, driver, and violation).

#### 5.2.2 Encrypting Violation Incidences

For encryption, each violation image is converted into a byte array; then, an XOR operation is applied on each value of the byte array using a predefined encryption key (2-bit key). After performing the XOR operation on each byte array value, the encrypted image is written to a new image file in a jpeg format. However, the driver won't be able to read the encrypted image or decrypt it. Only the ridesharing agent has the predefined key and can decrypt the image. The encryption key acts as a password to encrypt and decrypt the violation image.

#### 5.2.3 The Ridesharing Smart Phone

We ran experiments on two different smartphones with different specifications to monitor the performance of the proposed system on different hardware specifications. The first smartphone used during experiments was a Samsung s10 + phone running on Android. And the second phone is an LG V30+ also running on Android.
The system was built upon the starter code offered by android for object detection using *TensorFlowLite*[4]. The system added the custom trained Mobile SSD network. The code was adjusted to detect the three classes of driver, passenger, and violation. We added a *text-to-speech* module to send violation warnings. In addition, we added the encryption-decryption module for securing violation images.

#### 5.2.4 Results

These are some screenshots from the ridesharing violation application running on a Samsung s10 + phone. The application can detect three classes. A detecting bounding box surrounds each detected class. The number on the right top of the bounding box is the network confidence score in making this prediction. Examples of detected violations by the *SAFEMYRIDES* are shown in Fig. 4. a, 2. b. The confidence score of the DL network measures how confident is the network assigning an object a specific class.

## 6 Evaluation

The evaluation will be performed on different levels, as shown in table 3. The evaluation of the trained Mobile SSD network will be functional. The network is evaluated before and after it is compressed and quantized. Functional evaluation metrics include testing and training accuracy. We also traced the network testing and validation loss. The system itself is performed through controlled experiments where evaluation relativeness is absolute (e.g., test if the artifact achieves its goal?). Then, another evaluation is performed by conducting a comparison of the artifact performance on a different device of different hardware/software specifications (Cleven, Gubler, & Hüner, 2009). The stages of evaluation performed are summarized in table 3.

### 6.1 Evaluating the Trained Mobile SSD Network

#### 6.1.1 Training loss/ Validation Loss

There are two critical measures to evaluate the Mobile SSD: the training loss and the validation loss. The loss is the cost function used while training that needs to be minimized. At the end of every 100 training steps, the network is

---

[4] [GitHub Code](#)

evaluated for the loss in training and validation. As long as the loss decreases, that means that the network is trained correctly, and it can generalize well (Jancsary, & Rother, 2012). It is expected for the validation loss to be slightly less than the training loss. If the gap between training loss and validation loss increases, this indicates network overfitting. Across the 100K steps of training iterations, the training loss vs. validation loss is plotted as shown in fig. 4. The training and validation loss start high and start decreasing gradually. The gap between the training and validation loss is tight. However, at some steps, the validation loss is less than the training loss, which indicates that the validation samples are easier to detect than training samples. The iteration with the least loss is selected in step 96800 with a validation loss of 1.0199343. Then validation loss starts increasing afterward, which means that the network overfits. The MAP (Mean Average Precision) is another metric used in measuring the accuracy of object detecting networks. The average precision computes the average precision value for the recall values over 0 to 1 using IoU (Intersection over the union) (Henderson & Ferrari, 2016). the MAP was calculated at the end of each 10000 training steps at IoU=0.5 the maximum MAP achieved was `0.3816 at step 55867, and DetectionBoxes_Recall/AR@10 = 0.4333.`

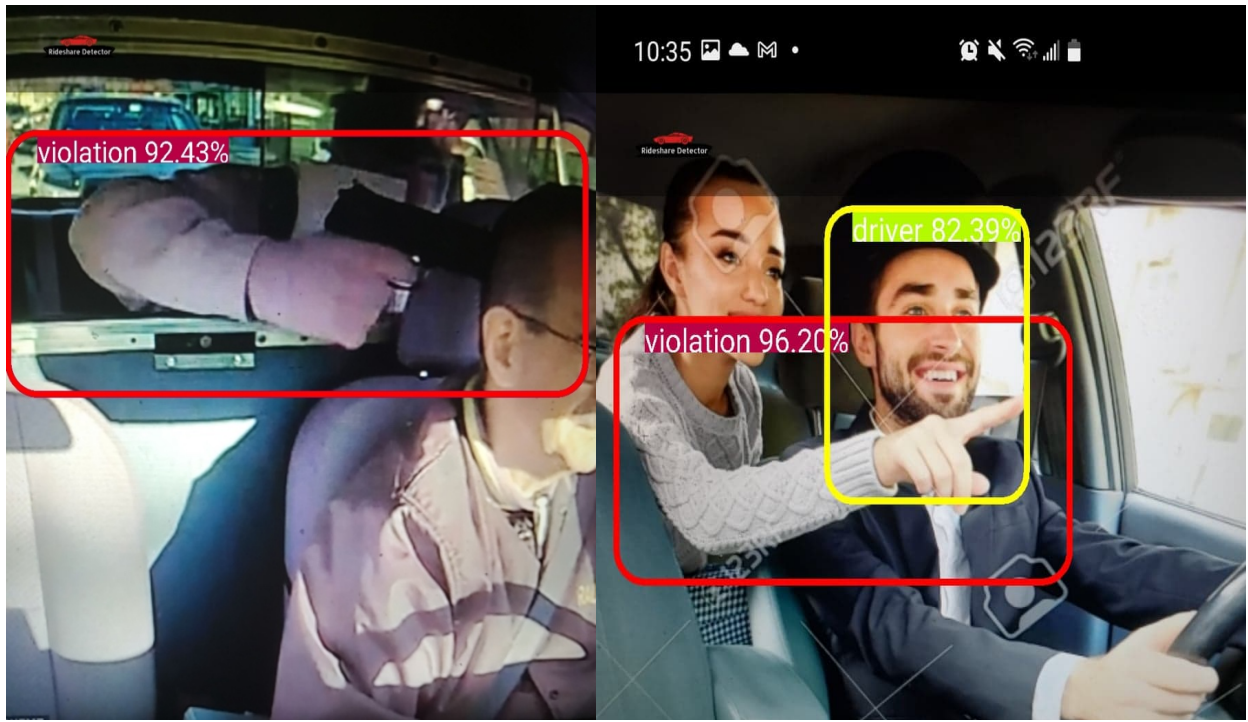

Fig. 4: a,b  a Violation of Sitting Next to The Driver, Interacting Physically with the Driver.

Table 3: Evaluation Stages with Involved Components and Criteria.

| Artifact | Evaluation Method | Assessed Criteria | Level of Evaluation | Form of Evaluation | Relativeness of Evaluation |
|---|---|---|---|---|---|
| Trained Mobile SSD | Reporting the functional performance of the trained network | Training accuracy Loss/validation accuracy Loss, | Instantiation | Quantitative / measured | Absolute |
| Proposed System | Demonstration of the performance of the artifact with several real examples | Goal, Accuracy, Latency | Instantiation / Real examples | Quantitative / measured | Absolute |
| Proposed System | Observing the differences between performance on different devices | Comparison between accuracy, latency efficiency on different devices | Instantiation | Quantitative / measured | Relative |

### 6.1.2 Training and Validation Accuracy

We ran an inference test on the trained network before we performed optimization: compression and quantization to study the effect of the optimization on accuracy and latency. The full-sized mobile SSD model sized 245 MB with a 32-bit inference parameter. The highest training accuracy achieved is 0.94546, and the highest validation accuracy is 0.93123. The average time spent by Google Collaboratory for inference testing is 3.56 seconds.

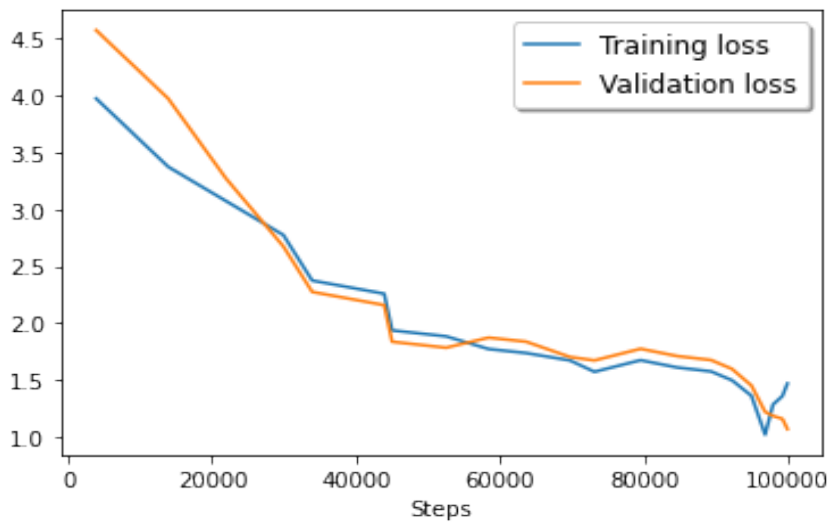

Fig. 5: Training loss Vs. Validation Loss of the Trained Mobile SSD.

## 6.2 Evaluating the SAFEMYRIDES Decision Support System

Firstly, the entire SAFEMYRIDES system was evaluated using a controlled simulated environment where the system was applied in a real ridesharing setting. The first experiments were running on the Samsung S10+ phone. The driver/passenger performed different violation scenarios such as: interacting physically with the driver, violence, and sitting next to the driver. Examples of detected violations are shown in Fig. 6 a, b.

### 6.2.1 Latency

The average system *Latency* is also measuring the system responsiveness that affects service quality and customer satisfaction. Latency would be represented by the average system response time to detect a violation. However, the response time is affected by the surrounding conditions. For example: the response time to detect a violation in daylight is different than the response time at night. After setting the minimum required confidence level to 0.80 (the confidence score of the Mobile SSD that the detected incidence is a violation), the average response time to violations detection during daylight is 450 ms while at night is 790 ms. However, this average response time is higher to other systems used in ridesharing media processing, as shown in table 4 and fig. 6.

Table 4: A Comparison Between SAFEMYRIDES and Other Proposed Systems.

| Research | Average Latency | Media type |
|---|---|---|
| L. Liu et al. (2018) | 1273ms | Sound |
| L. Wang et al. (2019) | 8345 ms | A 60-second video data |
| Long et al. (2017) | 2234 ms. | Compressed Videos. Latency (model processing time + network transmission time). |
| Ran et al. (2018) | 32100 ms. | video chunk is set as S= 1MB, 15 edge nodes |
| SAFEMYRIDES | 620 ms. | Scene view stream |

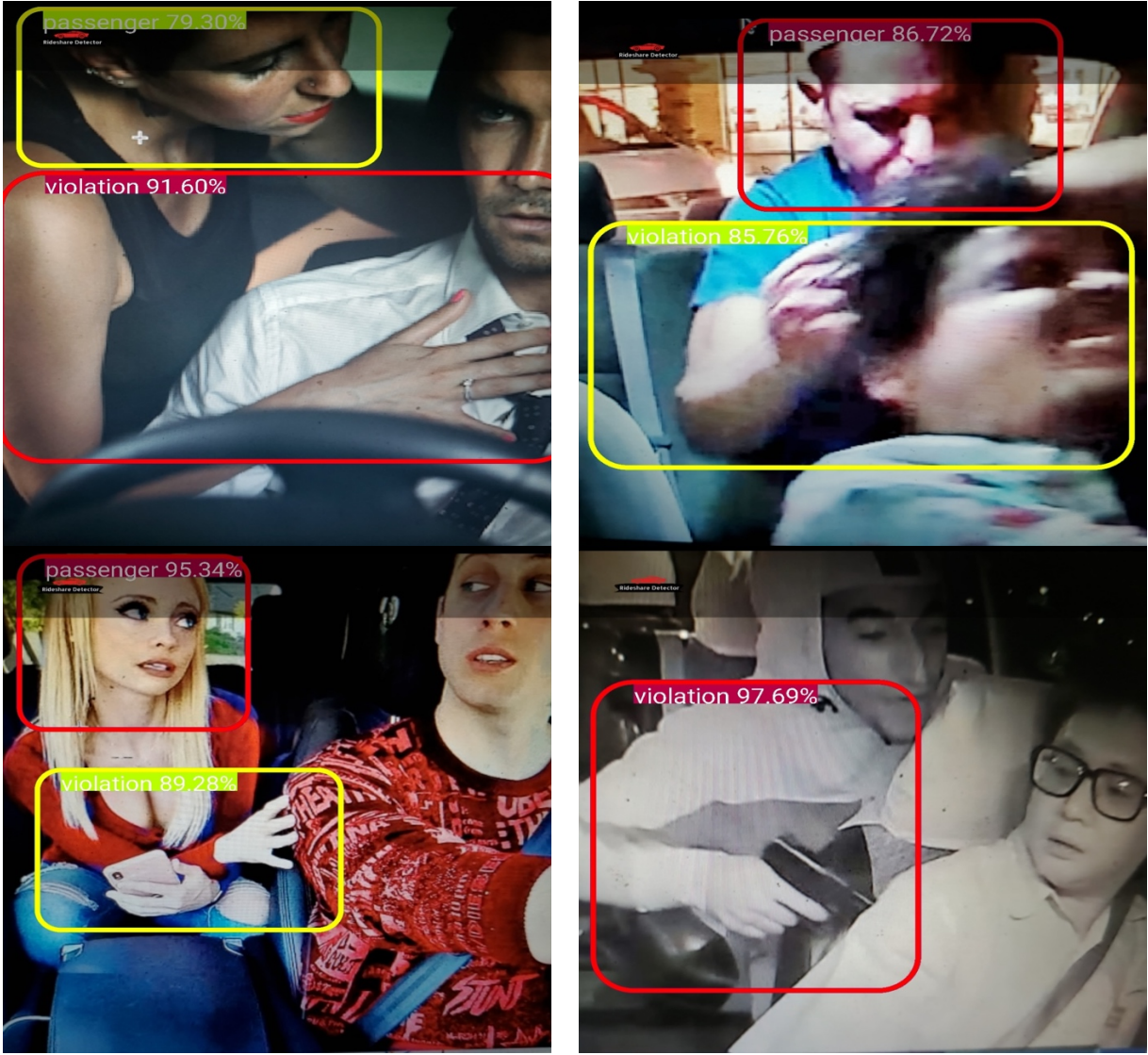

Fig. 6: Detected Violations using Different Violations: Violence and Physical Interaction.

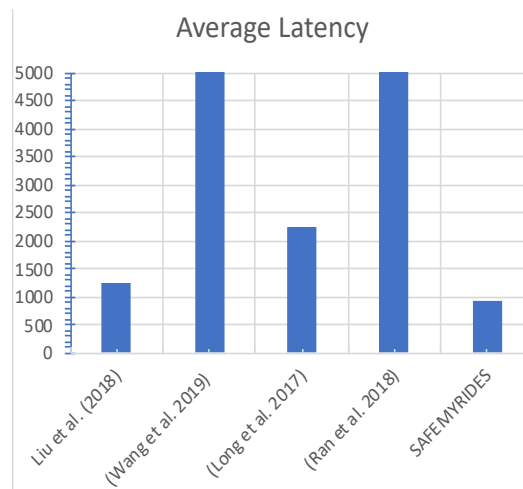

Fig. 7 : Average Latency of the Proposed System Versus Previously Developed Systems.

### 6.2.2 Efficiency

In terms of the resources used by the system, the system used 29.09 MB of the phone's internal storage. The average size of an encrypted media image is 120kB. Energy consumed of running the system for 15 minutes at a rate of 17 m (milliamp), or 79.8 mAH (milliamp hour). Memory used by the system is 104 MB Which means it is consuming less than 1% of the smartphone RAM of 8 GB.

### 6.2.3 The System Accuracy and Confusion Matrix

In this experiment, we are evaluating the optimized trained network running on the Android S10 plus. The optimized network size is 27 MB, almost 10x less than the actual model size, and runs at 8-bit inference parameters. To get an in-depth analysis of how the optimized network performed, a confusion matrix was created to analyze the accuracy, precision, and recall of the front and rear cameras as in table 5 and table 6. The confusion matrix was built during a simulated session during the day, trying 51 different normal and violations poses. The front camera and the rear camera were used to detect a violation. The rear cameras of the Samsung S10 plus are 12MP, 12 MP, and 16 MP ultra-wide module cameras. The front-facing cameras are 8MP, 10 MP f/1.9 selfie cameras.

**Precision:** It tells what fraction of predictions as a positive class were genuinely positive. To calculate precision, use the following formula: $TP/(TP + FP)$ or 0.9111 for the rear camera and 0.906 for the front camera according to the confusion matrix.

**Recall:** also known as Sensitivity, Probability of Detection. It tells what fraction of all positive samples were correctly predicted as positive by the classifier. To calculate Recall, use the following formula: $TP/(TP + FN)$ or 0.976 for the rear camera and 0.928 for the front camera.

**Accuracy:** is the measure of what fraction of violation incidences were correctly detected by the system; the accuracy formula is $Accuracy = (TP + TN)/(P + N)$. Based on the formula, the system's accuracy in detecting correct violation incidences is 46/51=0.9019 for the front camera and 0.862.

Table 5: Confusion Matrix of The Proposed System (Rear Camera).

| True/Detected | True | False |
| --- | --- | --- |
| True | 41 (TP) | 1 (FN) |
| False | 4 (FP) | 5 (TN) |

Table 6: Confusion Matrix of the proposed system (Front camera).

| True/Detected | True | False |
| --- | --- | --- |
| True | 39 (TP) | 3 (FN) |
| False | 4 (FP) | 5 (TN) |

### 6.2.4 Privacy

Surveillance systems have raised apprehension about their threat to individuals' privacy rights (Cheung, Venkatesh, Paruchuri, Zhao, & Nguyen, 2009). Privacy protection is a core objective in the proposed system through ensuring transparency, authorization, and encryption. Transparency means that both the driver and the passenger are aware that the ridesharing session is monitored. A warning of "Violation Detected" is declared when the system detects a violation. Media surveillance is privacy-intrusive because it allows the observation of certain information that is considered privacy intrusive such as a person's identity or characteristics (age, race, gender) (Senior, 2009) . Therefore, authorization of users is required to monitor ridesharing sessions. Moreover, negotiation of privacy preference (Consortium, 2002) or flexible authorization is offered when users prefer monitoring at night sessions or in remote destinations.

Privacy protection technologies have focused mainly on different visual obfuscation techniques. Encryption of violation incidences is applied to protect sensitive personal information from being saved and transferred securely (Diffie & Landau, 2010) . In the proposed system, an XOR operation with a (2 bit) encryption key is applied for every detected violation incidence and saved in a secured folder in the ridesharing phone. The detected violation incidences are only shared with ridesharing agents or official legal authorities.

## 6.3 Evaluating SAFEMYRIDES on Different Devices

In this section, the SAFEMYRIDES is applied to a different device to notice the discrepancy in the system performance with different hardware specifications. In the previous section, the system was applied to a Samsung S10 plus device. In this section, the system will be applied to an LG V30 plus phone to compare and contrast the system's performance on different hardware settings.

LG V30 plus has only two rear cameras (16 MP and 12 MP) and one 5MP front camera compared to three rear cameras and two front cameras in the Samsung S 10 plus. The LG V30 Adreno 540 GPU is compared to an Adreno 640 GPU in S10+. Also, the S10 + has a Li-Ion 4100 mAh Battery, and 8 GB RAM is compared to Li-Po 3300 mAh and 4 GB in LG V30. Table 7 and Fig. 7 compares the performance of SAFEMYRIDES on the two devices in terms of average latency, average accuracy, and efficiency (the resources used) for running the system for 15 minutes. The comparison is performed using the rear cameras because of higher resolution and night shots adjustment features.

While the Samsung S10 + achieved higher performance and less latency than The LG V30, the resources consumed (battery and RAM were higher). Samsung S10 + has three cameras which consumed more battery power and RAM in return. There is almost a 4 % difference in the accuracy between the LG V30 and Samsung S10 + and more than 350 ms difference in Latency.

Table 7: Performance Comparison for Applying SAFEMYRIDES on Different Devices.

| Metric/ Device | LG V30 (Rear Camera) | Samsung S 10 plus (Rear Camera) |
|---|---|---|
| Accuracy | 0.8845 | 0.9019 |
| Latency | 1297 ms | 920 ms. |
| RAM | 89 MB | 104 MB |
| Battery Consumption | 48.7 mAH (milliamp hour) | 79.8 mAH (milliamp hour). |
| Internal Storage | 29.8 MB | 28.09 MB |
| Average CPU | 0.72 | 0.64 |

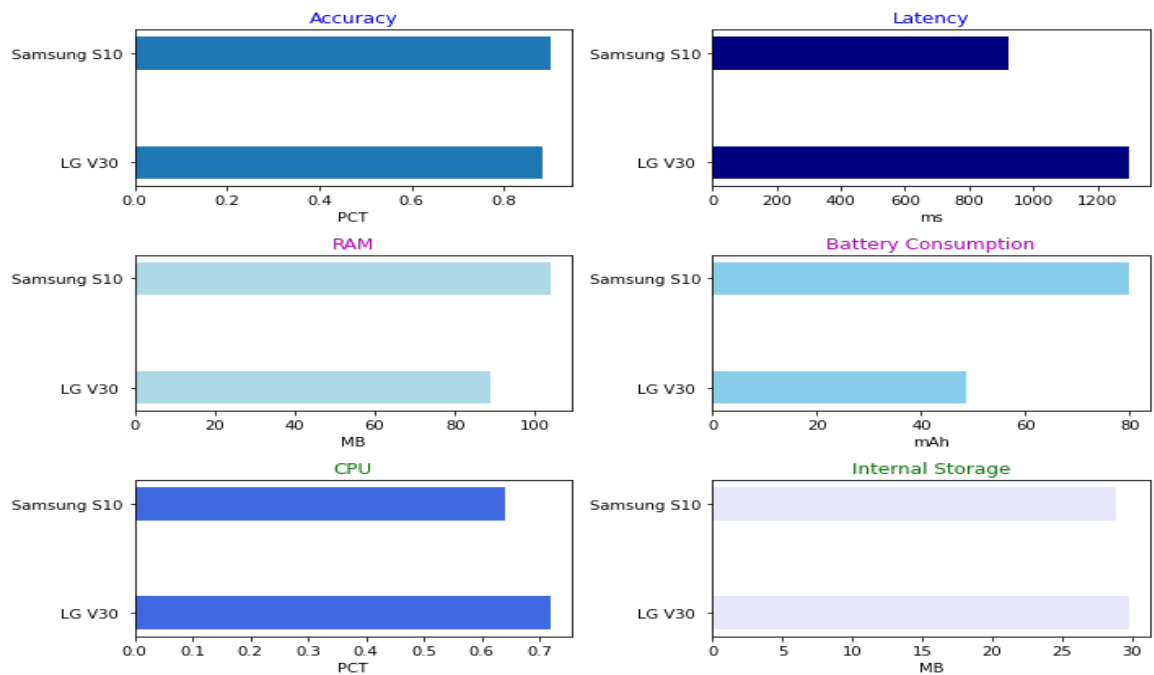

Fig. 8: Performance Comparison between LG and Samsung using Rear Camera.

# 7     Discussion and Implications

While most of current edge computing approaches focus on optimizing processing and data transfer on edge units, little research focus on empowering IoT to decrease dependency on task offloading. So, we developed a decentralized-control edge computing model where computation and decision making are moved to the IoT level. Moving the computation to IoT (onsite) will enhance *efficiency, responsiveness, accuracy, and privacy* issues of many IoT applications. The decentralized-control model is fueled by technologies such as *TF lite* and IoT optimization techniques which empower IoTs to run cognitive tasks at humble consumption of resources. As a proof of concept, we build a scene-aware ridesharing monitoring system to detect violations at runtime with an average latency of 620 ms. The system was evaluated technically and functionally using controlled experiments by stimulating different violation poses in real ridesharing settings.

The system was designed to ensure *safety and privacy*. So, violation incidences are saved in hidden folders in an encrypted format with an average file size of 120KB. The system responsiveness (latency) was comapred against previously developed systems. The system achieved the least latency and reduced the cost of transferring images for processing on edge nodes. Instead, the system detects violations locally on the driver's smartphone. Given the limited number of training samples, the system achieved a 90.1% accuracy rate for detecting 51 different violations poses with a 94% sensitivity rate. On the other hand, we can notice the significant difference in average inference time (delay) between the full network of 3.56 seconds and the optimized network of 920 ms which gives optimized DL networks an additional advantage to be applied for real time response.

When the system was applied to different devices, a variance in performance was detected which affected the detection accuracy and latency. The higher is the resolution of the phone's camera, the better is the accuracy. Moreover, the higher is the GPU/CPU, the less is the system latency. In addition, detection using rear cameras is better than front cameras. Moreover, we noticed there is not much difference in detection accuracy during the day, but there is a difference in detection accuracy at night. In general, the lower detection accuracy at night calls for increasing the number of night training samples. Moreover, since most smartphone cameras don't have night vision, the camera ability to balance colors of night images affects detection accuracy. Also, the false-positive rate calls for training the network with more regular ridesharing poses. The system is efficient to use in terms of storage (29 MB) and memory consumption (less than 1% of active memory (RAM) consumption).

# 8     Limitations and Future Work

The implementation of the ridesharing monitoring system was performed as a proof of concept for the decentralized control edge computing model. The size of the dataset used was limited. Increasing the training and validation samples will positively affect the accuracy and the latency of predictions. In addition, the encryption of violation incidences was performed using a simple XOR function. More sophisticated encryption methods using a 128-Bit key are preferred. Also, additional privacy preservation mechanisms such as video inpainting and data-hiding schemes will add extra confirmation on user privacy (Cheung et al., 2009). Since the system achieved lower accuracy and decreased sensitivity in detecting violations during the night, the deep learning network needs to be trained with more night samples, where poor light obstructs definite detection of violations.

Future research on the optimization of camera usage is required in terms of battery consumption as the energy consumption of cameras is considered above average, assuming that a ridesharing driver will be using the system for hours. Moreover, the system should be deployed in a real setting to address other intangibles quality measures such as reliability, understanding, and courtesy.

Other future work includes implementing the ridesharing monitoring system on other platforms such as IOS and Raspberry PI. The system should be tested on different platforms for benchmarking. Benchmarking will help determine the minimum required for a smartphone to run the system. While the system used compression and quantization techniques such as optimization methods for the Mobile SSD, applying other optimization techniques such as pruning might enhance the optimized network accuracy.

The proposed system is trying to solve the *safety and privacy* issues related to ridesharing, and it has several implications to practice. Firstly, the system allows local ridesharing monitoring on the driver's smartphone, saving the cost of an additional device for monitoring. Having the monitoring system running on the same device used in the ridesharing will ensure efficiency and privacy. Secondly, the system uses no network communication while detecting violations and limits the network communication to reporting the violation incidences to the ridesharing agent.

Minimizing network communication ensures *privacy, security, and efficiency*. So, the system could be used where poor network connections exist. Thirdly, encryption of ridesharing incidences ensures privacy and security. So, the passenger/driver data are only shared directly with the ridesharing agent in case of detected violation.

This research contributes to the *Information System* research by applying the concept of decentralized control where an IoT (smartphone) can independently detect violations and make decisions. The research is one of the early attempts to apply different optimization methods to deep learning networks to run efficiently on an IoT. The research sheds light on the importance of limiting cloud/edge computation and limiting network communications while enhancing the decisions making process and analytics of the IoTs.

## 9 Conclusion

This research is introducing a decentralized-control edge computing model where computation and decision making is moved to the IoT level. The model addresses the *efficiency, responsiveness, and privacy* issues that current edge computing models still contemplate. Although battery powered, IoTs could run cognitive tasks at reasonable consumption of resources using different optimization techniques. We applied the model to build a scene-aware ridesharing monitoring system to detect violations at runtime. The system uses an optimized deep learning model on IoTs that can locally detect violations in ridesharing sessions on the driver's smartphone. The system records violations incidents only and saves them in an encrypted format. So, the system addresses the *safety* issue without violating *privacy*. On the technical level, the decentralized control edge model is reducing communication cost and latency, ensure privacy and less prone to data transfer security risks.

## 10 Declaration of interests

The authors declare the following financial interests/personal relationships which may be considered as potential competing interests. This research was supported in part by 2019 Small Grants Award from Virginia Commonwealth School of Business Information Systems Department.